\documentclass{PoS}
\usepackage{amssymb}
\usepackage{amsmath}
\usepackage{textcomp}
\usepackage[utf8]{inputenc}
\usepackage{upgreek}
\usepackage{lineno}

\def\bi{\begin{itemize}}
\def\ei{\end{itemize}}

\title{Status of cosmic-ray antideuteron searches}

\ShortTitle{Status of cosmic-ray antideuteron searches}

\author{\speaker{P. von Doetinchem}, R. Pereira\\
        University of Hawai'i at M$\bar{\text{a}}$noa\\
        E-mail: \email{philipvd@hawaii.edu}}
\author{T. Aramaki, C.J. Hailey\\
Columbia University}       
        
\author{S. Boggs\\
University of California at Berkeley}

\author{S.~Bufalino\\INFN Sezione di Torino}

\author{L.~Dal, A.~Raklev\\University of Oslo}

\author{F.~Donato, N.~Fornengo, A.~Vittino\\INFN Sezione di Torino, University of Torino}

\author{H. Fuke\\
Institute of Space and Astronautical Science, Japan Aerospace Exploration Agency}

\author{M.~Grefe\\Universit\"at Hamburg}

\author{B. Hamilton\\University of Maryland}

\author{A.~Ibarra, S.~Wild\\Technische Universit\"at M\"unchen}

\author{J.~Mitchell, M.~Sasaki\\NASA/Goddard Space Flight Center}

\author{S.I. Mognet, R.A. Ong\\
University of California at Los Angeles}

\author{K. Perez\\
Haverford College}

\author{A.~Putze\\LAPTh, Universite Savoie Mont Blanc, LAPP, Universite Savoie Mont Blanc}

\author{P.~Salati\\LAPTh, Universite Savoie Mont Blanc}

\author{G.~Tarle\\University of Michigan}

\author{A.~Urbano\\SISSA - International School for Advanced Studies}

\author{W.~Xue\\Massachusetts Institute of Technology}

\author{K.~Yoshimura\\High Energy Accelerator Research Organization (KEK)}

\abstract{The precise measurement of cosmic-ray antiparticles serves as important means for identifying the nature of dark matter. Recent years showed that identifying the nature of dark matter with cosmic-ray positrons and higher energy antiprotons is difficult, and has lead to a significantly increased interest in cosmic-ray antideuteron searches. Antideuterons may also be generated in dark matter annihilations or decays, offering a potential breakthrough in unexplored phase space for dark matter.  Low-energy antideuterons are an important approach because the flux from dark matter interactions exceeds the background flux by more than two orders of magnitude in the low-energy range for a wide variety of models. This review is based on the ``$\bar {\text d}$14 -- 1$^{\text{st}}$ dedicated cosmic-ray antideuteron workshop'', which brought together theorists and experimentalists in the field to discuss the current status, perspectives, and challenges for cosmic-ray antideuteron searches and discusses the motivation for antideuteron searches, the theoretical and experimental uncertainties of antideuteron production and propagation in our Galaxy, as well as give an experimental cosmic-ray antideuteron search status update. This report is a condensed summary of the article ``Review of the theoretical and experimental status of dark matter identification with cosmic-ray antideuteron''~\cite{2015arXiv150507785A}.}

\FullConference{The 34th International Cosmic Ray Conference,\\
		30 July- 6 August, 2015\\
		The Hague, The Netherlands}

\begin{document}

\section{Indirect dark matter search}

The existence of dark matter is established on very different length scales from galaxies to galaxy clusters to the cosmic microwave background~\cite{darkmatter}. If dark matter was in thermal equilibrium in the early universe, and froze out when the temperature dropped due to expansion, it is a natural assumption that dark matter particles are able to interact with each other and produce Standard Model particles. Indirect searches exploit possible kinematic differences between the production of cosmic rays through dark matter and standard astrophysical processes to identify dark matter signals. Cosmic-ray antiparticles without primary astrophysical sources are ideal candidates for an indirect dark matter search, but recent results show that accomplishing this task with positrons and antiprotons is challenging due to high levels of secondary/tertiary astrophysical background. However, the latest results of major cosmic-ray instruments (e.g., AMS-02~\cite{ams4}) for the positron fraction show evidence of a structure that might be interpreted as dark matter. Recently released AMS-02 antiproton-to-proton ratio data is inconclusive (e.g., \cite{2015arXiv150404276G,2015arXiv150501236K,2015arXiv150405937H}).

\subsection{Antideuterons \label{s-2}}

Antideuterons may be generated in dark matter annihilations or decays, offering a potential breakthrough in unexplored phase space for dark matter. The unique strength of a search for low-energy antideuterons lies in the ultra-low astrophysical background~\cite{2015arXiv150507785A}. 
The dominant conventional sources for secondary (background) antideuteron production are cosmic-ray protons or antiprotons interacting with the interstellar medium~\cite{Duperray:2005si}. However, the high threshold energy for antideuteron production and the steep energy spectrum of cosmic rays mean there are fewer particles with sufficient energy to produce secondary antideuterons, and those that are produced have relatively large kinetic energy.

\subsection{Discovery potential of dark matter searches with antideuterons\label{s-2}}

\begin{figure}
\centering
\includegraphics[height=0.4\linewidth]{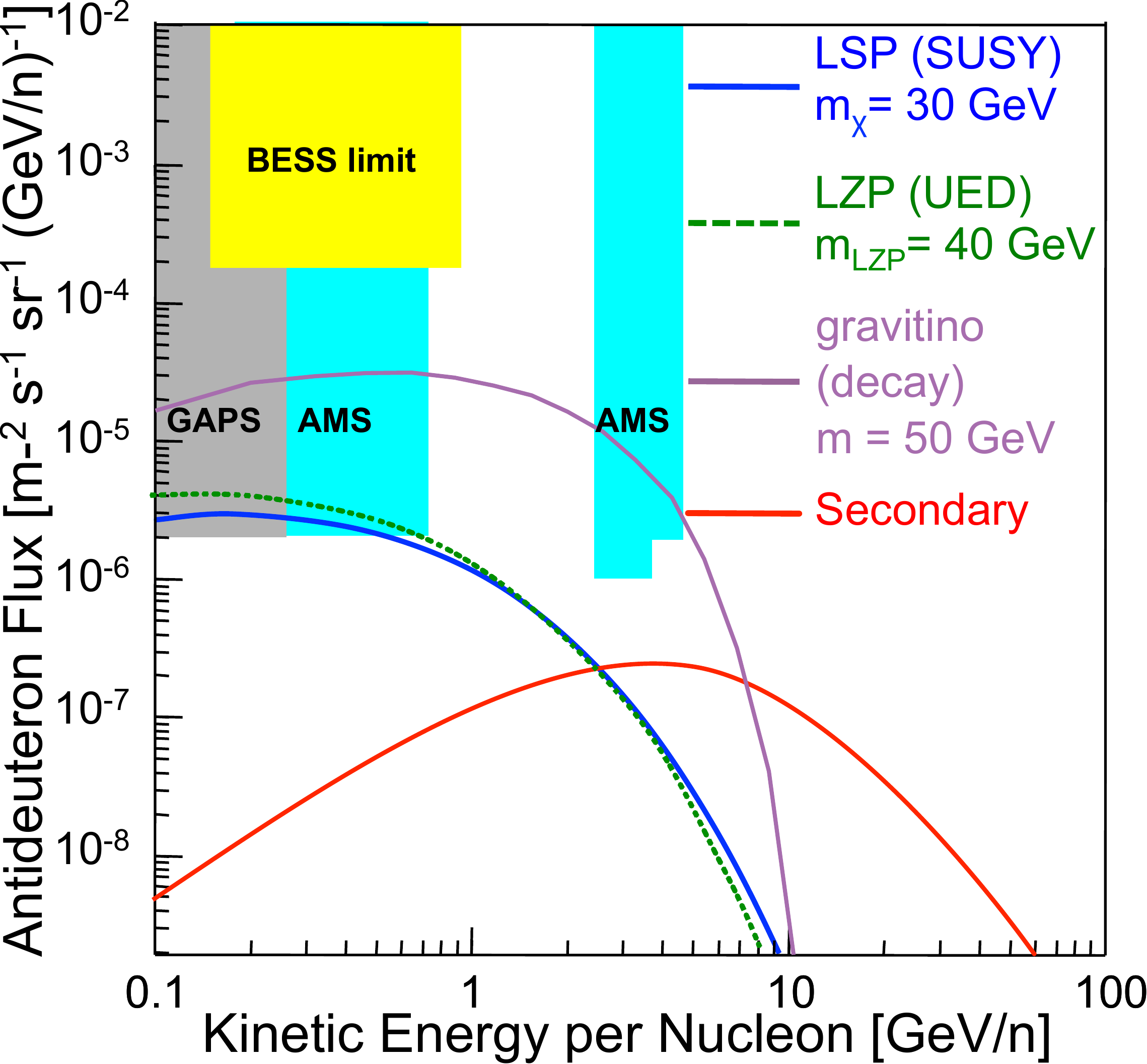}
\hspace{.1\linewidth}
\includegraphics[height=0.4\linewidth]{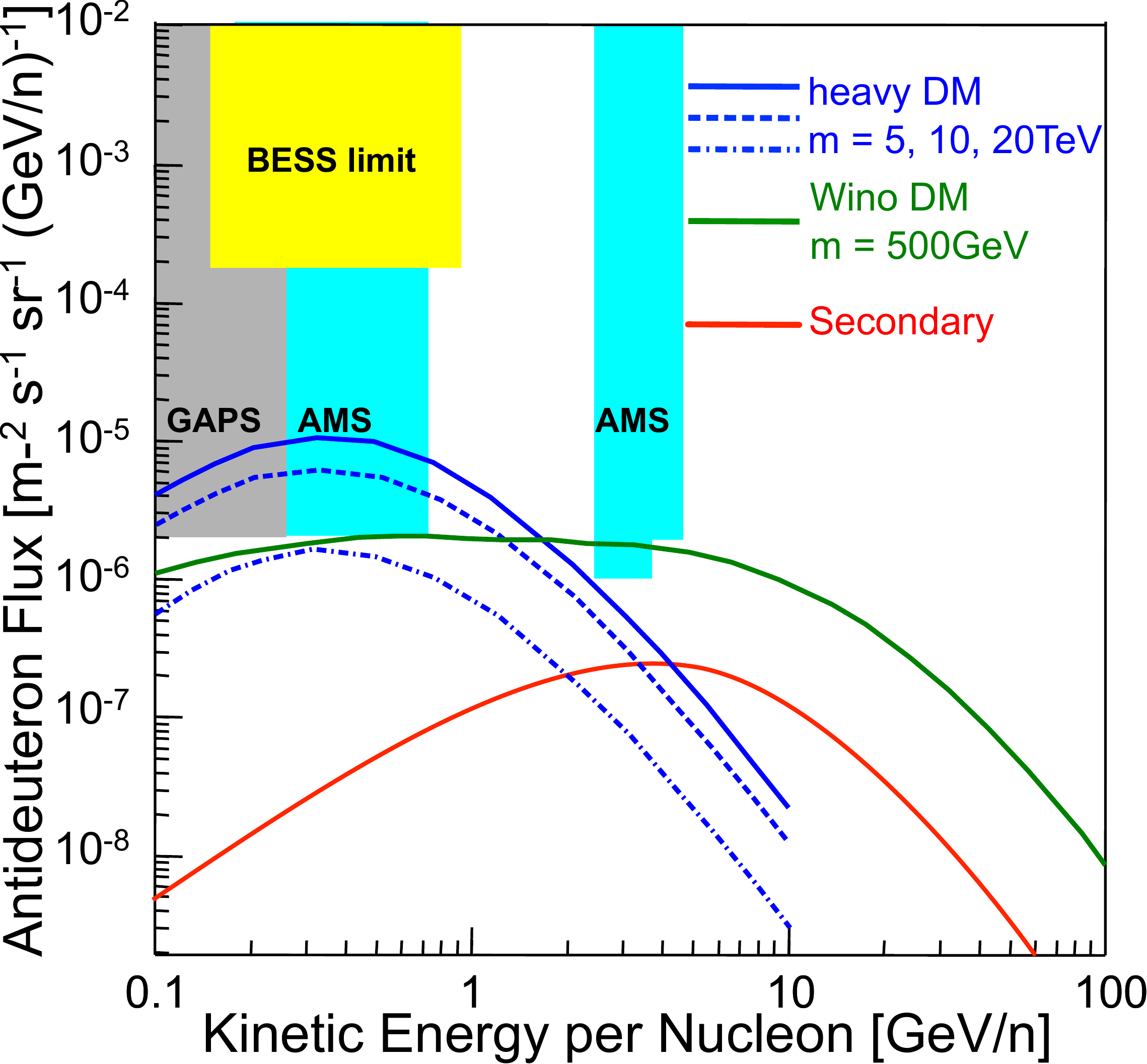}
\caption{\textit{Left)} Antideuteron limits from BESS \cite{Fuke:2005it}, predicted antideuteron fluxes from different models \cite{Baer:2005tw,Donato:2008yx,Dal:2014nda,Ibarra2013a}, sensitivities for AMS-02 for 5\,years \cite{ams02dbaricrc2007} and the planned GAPS experiments after three 35-day flights \cite{gaps,Aramaki2015}. \textit{Right)} Predicted antideuteron flux for annihilation of dark matter with $m_{\text{DM}} =$~5, 10, 20~TeV~\cite{Brauninger:2009pe} (blue lines, top to bottom) into $b\bar b$ and from pure-Wino dark matter~\cite{Hryczuk:2014hpa} (solid green line). \label{f-dmdbar}}
\end{figure}

Many dark matter models with masses from $\mathcal{O}(1\text{\,GeV})$ to $\mathcal{O}(1\text{\,TeV})$ potentially produce an antideuteron flux that is within the reach of currently operating or planned experiments, AMS-02 and GAPS (Sec.~\ref{s-5}). The left side of Fig.~\ref{f-dmdbar} shows the antideuteron flux expected from three benchmark dark matter scenarios. These dark matter candidates include a lightest supersymmetric particle (LSP) neutralino from the minimal supersymmetric standard model (MSSM), a 5D warped GUT Dirac neutrino (LZP), and an LSP gravitino. The expected secondary/tertiary background~\cite{Ibarra2013a} is also shown. This figure reveals why low-energy antideuterons are such an important approach: the flux from a wide range of viable dark matter models exceeds the background flux by more than two orders of magnitude in the energy range below 0.25\,GeV/$n$, and by more than an order of magnitude up to 1\,GeV/$n$. However, antideuterons are not only sensitive to models with dark matter masses in the 10--100\,GeV range, but also to heavy dark matter models with masses 0.5--20\,TeV, motivated by the positron fraction excess~\cite{Brauninger:2009pe}. These models require the MAX propagation model (Sec.~\ref{s-trans}) and an enhanced annihilation cross section, such as provided by the Sommerfeld mechanism. The case for these multi-TeV mass particles annihilating into $b\bar b$ is shown on the right side of Fig.~\ref{f-dmdbar}. The same figure also illustrates the case of heavy supersymmetric pure-Wino dark matter~\cite{Hryczuk:2014hpa}.

It is vital to note that every process that produces antideuterons will also produce a much larger flux of antiprotons and any prospective antideuteron signature from dark matter is constrained by antiprotons. However, detecting deviations from the astrophysical antiproton flux requires very high statistics. Therefore, antideuterons provide an additional search channel with very strongly suppressed astrophysical backgrounds compared to antiprotons, and  can also act as an essential probe to confirm or rule out potential deviations in the antiproton spectrum due to processes like dark matter annihilation or decay. On the other hand, a non-detection of a signal above background in the antiproton channel might also just be a consequence of experimental limitations and serves as additional motivation to pursue antideuteron searches.

\section{Uncertainties for cosmic-ray antideuteron searches}

The following discusses theoretical and experimental uncertainties for antideuteron searches. After antideuterons have formed, they propagate through our Galaxy to reach the solar system, where they can be deflected by the solar magnetic field or suffer adiabatic energy losses in the solar wind. Finally, antideuterons can be deflected away from balloon-borne or satellite detectors by the geomagnetic field and/or interact with Earth's atmosphere. It is important to point out that the antideuteron fluxes presented assume a conservative boost factor, due to dark matter clumps in the galactic halo, of $f=1$. However, a boost factor of $f=2$--3, as is consistent with current theoretical expectations, would increase dark matter fluxes by a factor $f$ over those discussed below~\cite{Baer:2005tw,Lavalle:1900wn}. Such a boost factor is only relevant for dark matter annihilation, as dark matter decay depends linearly on the dark matter density. In addition, the choice of the dark matter density distribution profile has only a small effect on the primary antideuteron flux and is not further discussed~\cite{Fornengo:2013osa}.

\subsection{Antideuteron production\label{s-3}}

\begin{figure}
\centering
\includegraphics[width=0.75\linewidth]{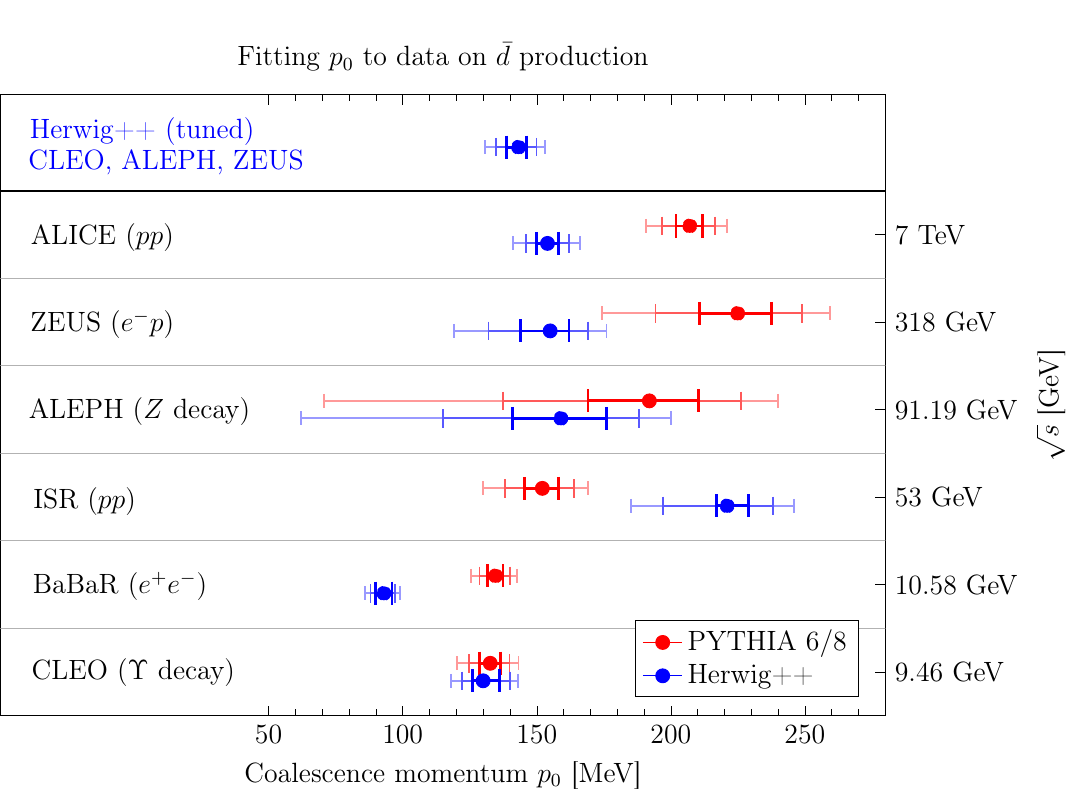}
\caption{Results of fitting the coalescence momentum $p_0$ to different data sets on antideuteron production, based on \cite{Ibarra:2012cc} and \cite{Dal:2014nda}.}
\label{fig:p0_determination}
\end{figure}

Understanding antideuteron production is crucial for the interpretation of cosmic-ray data, which impacts both the astrophysical background as well as the predictions from dark matter annihilations or decays. The fusion of an antiproton and an antineutron into an antideuteron is described by the coalescence model, which is based on the assumption that any (anti)nucleons within a sphere of radius $p_0$ in momentum space will coalesce to produce an (anti)nucleus. The coalescence momentum $p_0$ is a phenomenological quantity, and has to be determined through fits to experimental data~\cite{PhysRev.129.854}. While the assumption of isotropic and uncorrelated nucleon spectra might be a good approximation in low-energy or minimum bias nuclear interactions for which the model was made, 
these assumptions do not hold in relevant elementary particle interactions~\cite{Kadastik:2009ts}, such as dark matter annihilations or decay, and $pp$ collisions at low center-of-mass energies. The state of the art is to apply the coalescence condition to $\bar{p}\bar{n}$-pairs on a per-event basis in Monte Carlo events. However, different event generators yield different values of $p_0$ when compared to a particular experiment, indicating a substantial systematic uncertainty in the coalescence prediction (Fig.~\ref{fig:p0_determination}). It is currently an open question if the antideuteron production depends on the exact underlying process and on the available center-of-mass energy or if the generators need further refinement.

In order to make progress in the understanding of antideuteron formation and the prediction of the primary and secondary antideuteron fluxes, more experimental data and a better modeling of antideuteron formation is needed. Recently, new experimental data of the $B\!{\scriptstyle A}\!B\!{\scriptstyle A\! R}$~\cite{Lees:2014iub} and ALICE~\cite{Barile:2014wma} experiments have become available that can be used for the determination of the coalescence momentum, and further independent determinations of the coalescence momentum will likely also be possible in the near future. To reduce the systematic uncertainties and to correctly interpret a hypothetical future detection of cosmic antideuterons, an experiment measuring antideuteron production in $pp$ collisions at low center-of-mass energies, i.e.\ $\sqrt{s} \approx 10$\,GeV, would be of utmost importance. In addition, also studies of antideuteron production in processes like $p$-C are important for constraining instrumental backgrounds and reducing systematic effects. The operational fixed target experiment NA61/SHINE is ideally suited for these tasks~\cite{na61}.

\subsection{Transport in the galactic environment\label{s-trans}}

\begin{figure}
\centering
\includegraphics[height=0.4\linewidth]{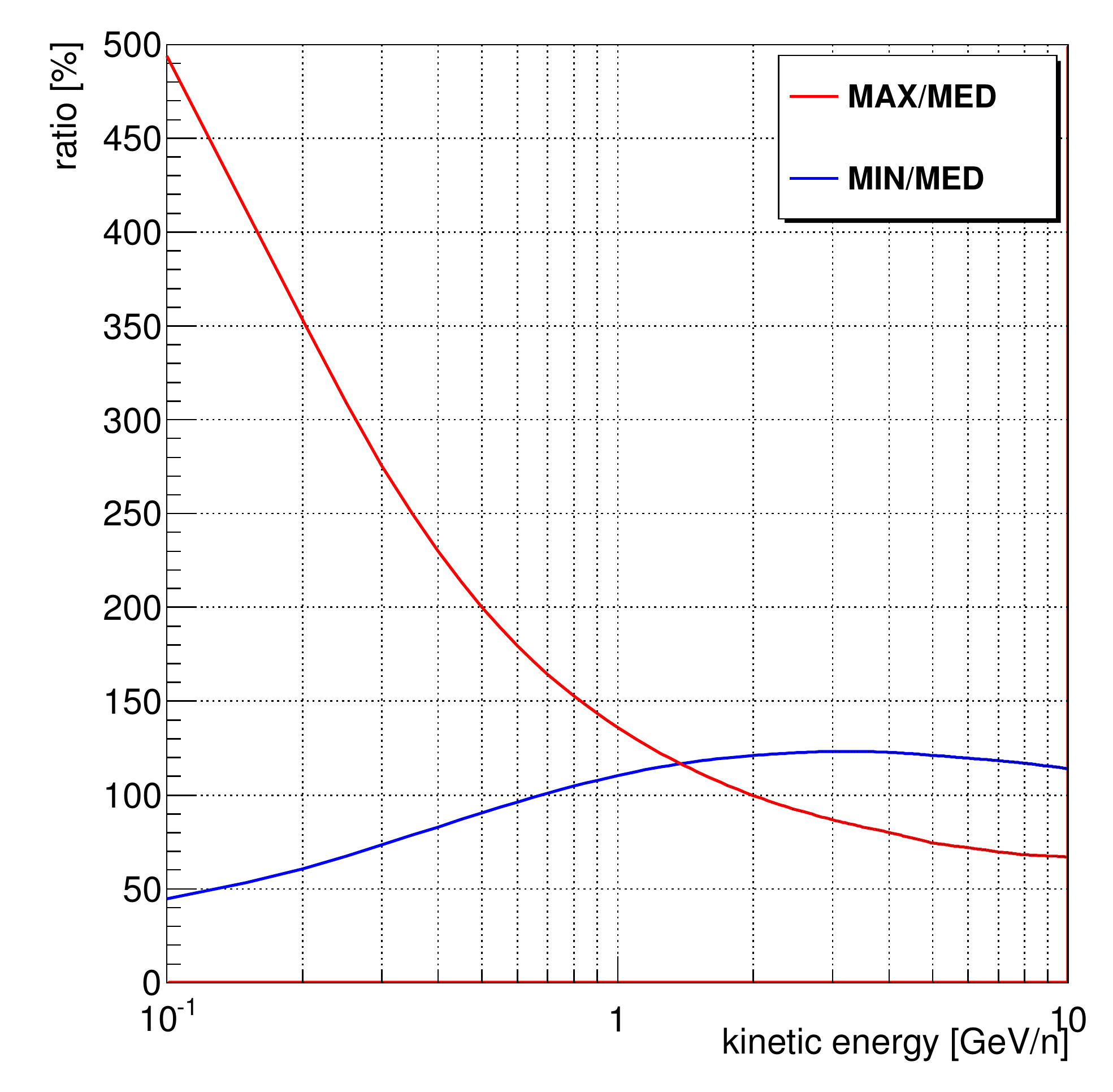}
\hspace{.1\linewidth}
\includegraphics[height=0.4\linewidth]{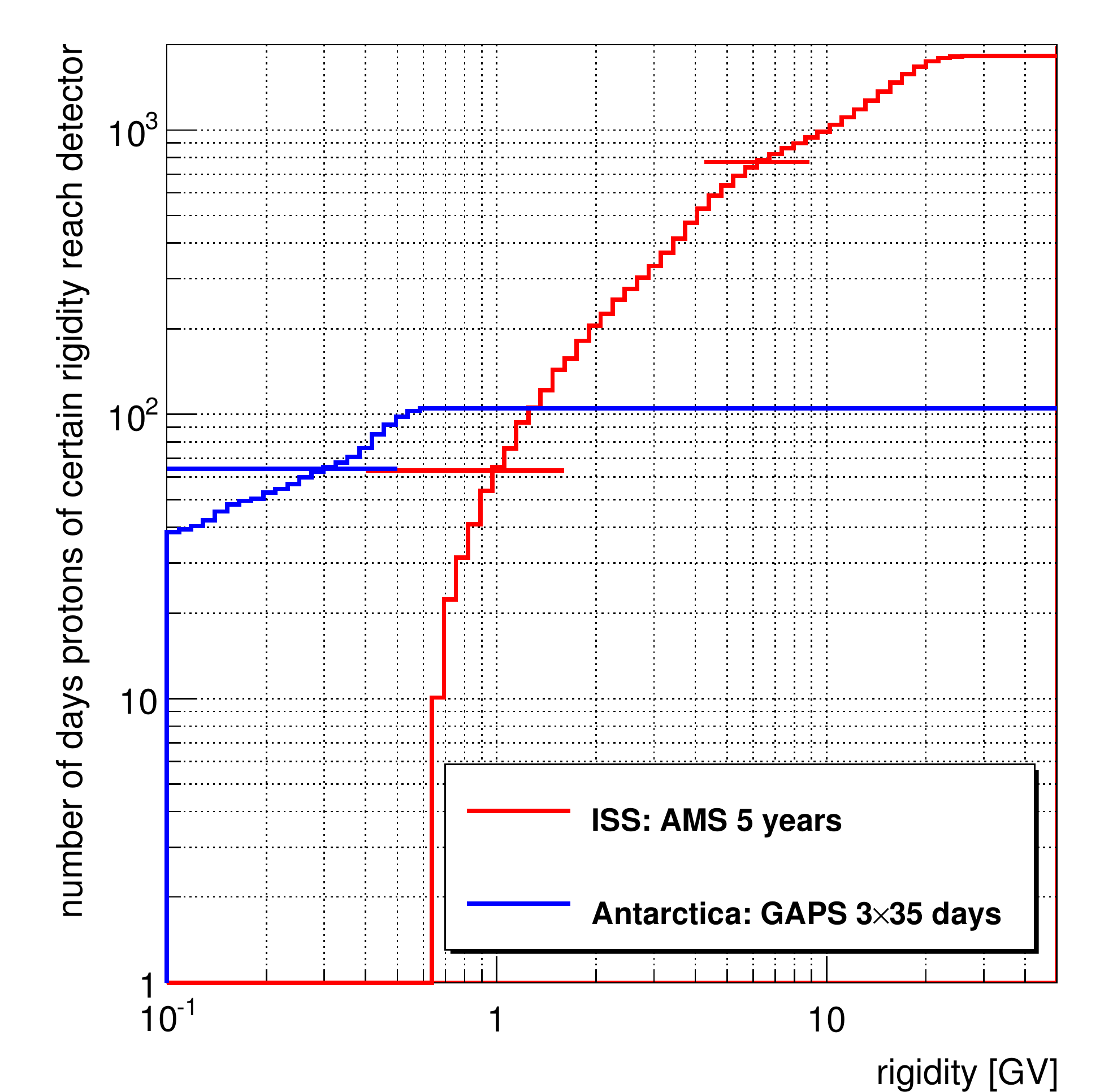}
\caption{\textit{Left)} Predictions of MIN and MAX propagation models for antideuteron top-of-atmosphere fluxes for dark matter with mass 100\,GeV annihilating into $b\bar{b}$ normalized to MED propagation model. \textit{Right)} Integrated measurement time above the geomagnetic cutoff for the AMS-02 and GAPS experiments. \label{fig:antid_propleft}}
\end{figure}

A full numerical treatment is generally required to solve the transport equation, as described, e.g., in~\cite{1998ApJ...509..212S}. However, (semi-)analytical solutions may be derived assuming a simplified description of the spatial dependence of some parameters. The two-zone diffusion model, based on the description of the Galaxy as a thin gaseous disk embedded in a thick diffusive halo, has been extensively studied (e.g.,~\cite{Maurin:2001sj}). In this framework, the details of galactic propagation depend entirely on the values that are assigned to the halo thickness $L$, convection velocity $V_{c}$, diffusion coefficient $K_0$ and exponent $\delta$, as well as the Alfv\'enic speed $V_a$. These parameters are determined by measurements of cosmic-ray observables such as the primary fluxes and secondary-to-primary ratios (in particular B/C). However, many sets of these parameters lead to the same B/C ratio and the same secondary antiproton flux within errors~\cite{2001ApJ...563..172D}. In particular, a large uncertainty for dark matter searches stems from the degeneracy between the normalization of the diffusion coefficient $K_0$ and the halo size $L$, where $L$ has a strong impact on dark matter signal predictions since it is proportional to the amount of cosmic rays induced by dark matter annihilation or decay within the diffusive Galactic Halo. As benchmark scenarios, the three sets of parameters labeled as MIN, MED and MAX defined in \cite{Donato:2003xg} are adopted. The values $L=1$ and 15\,kpc are used in the MIN and MAX models to bracket the theoretical propagation uncertainties, relying on the boron-to-carbon (B/C) analysis performed in~\cite{Maurin:2001sj}. The relative difference in the flux predicted by the MIN and MAX models is demonstrated in the left panel of Fig.~\ref{fig:antid_propleft} and spreads about one order of magnitude for low energies. Recent positron data exclude the MIN galactic propagation model~\cite{2014PhRvD..90h1301L}, which predicts the lowest antideuteron flux levels, and thus supports higher antideuteron flux predictions.

\subsection{Solar modulation}

When charged cosmic rays reach the heliosphere, they experience diffusion due to inhomogeneities in the solar magnetic field, begin drifting along the field lines, and suffer adiabatic energy loss in the solar wind. Different approaches have been discussed from the simple force-field approximation~\cite{gleeson-1967} to numerical simulations~\cite{Maccione:2012cu}. Compared to the uncertainties in the coalescence model and galactic propagation, the uncertainties for solar modulation are rather small ($\approx20$\%).

\subsection{Geomagnetic deflection}

The geomagnetic field is roughly described by a tilted dipole field, which provides the strongest charged particle shielding at the equator and the weakest at the poles. Backtracing charged particles through the geomagnetic field using the International Geomagnetic Reference Field and the Tsyganenko 2001 for the external magnetosphere \cite{JGRA:JGRA16156}), the right side of Fig.~\ref{fig:antid_propleft} compares the number of days that protons can reach AMS-02 (over five years of orbit) and GAPS (over three 35-day Antarctic flights) as a function of particle rigidity. As the geomagnetic cutoff for the Antarctic GAPS trajectory is generally much lower than that at the ISS, three 35-day GAPS flights show an exposure to low-energy particles that is comparable to five years of AMS-02 orbit.

\subsection{Atmospheric influence}\label{s-atmo}

The kinetic energy loss of cosmic-ray antideuterons in the atmosphere after reaching a typical balloon-experiment altitude of 37\,km above Antarctica is at the 10\%-level, and thus the measured kinetic energy can be corrected to top-of-the-atmosphere kinetic energies without introducing large systematic uncertainties. In addition, atmospheric antideuteron production is also important for the space-based AMS-02 experiment. Atmospherically produced antideuterons can upscatter into space and will create low-energy antideuterons at high geomagnetic cutoff locations.

\subsection{Antideuteron interactions in particle physics detectors}\label{s-geant}

One of the standard tools for studying particle interactions with detectors is Geant4. Until recently, Geant4 did not allow for the study of light antinucleus-nucleus interactions. The authors of~\cite{Galoyan:2012bh} added light antinuclei capabilities to Geant4 using the Glauber approach for the cross sections, the quark-gluon string model for annihilations and meson production, and the binary cascade for secondary interactions of low-energy mesons. This model was tested between 0.1\,GeV/$n$ and 1\,TeV/$n$, but needs further validation.

\section{Experiments for the detection of cosmic-ray antideuterons\label{s-5}}

\begin{figure}
\centering
\includegraphics[width=0.76\linewidth]{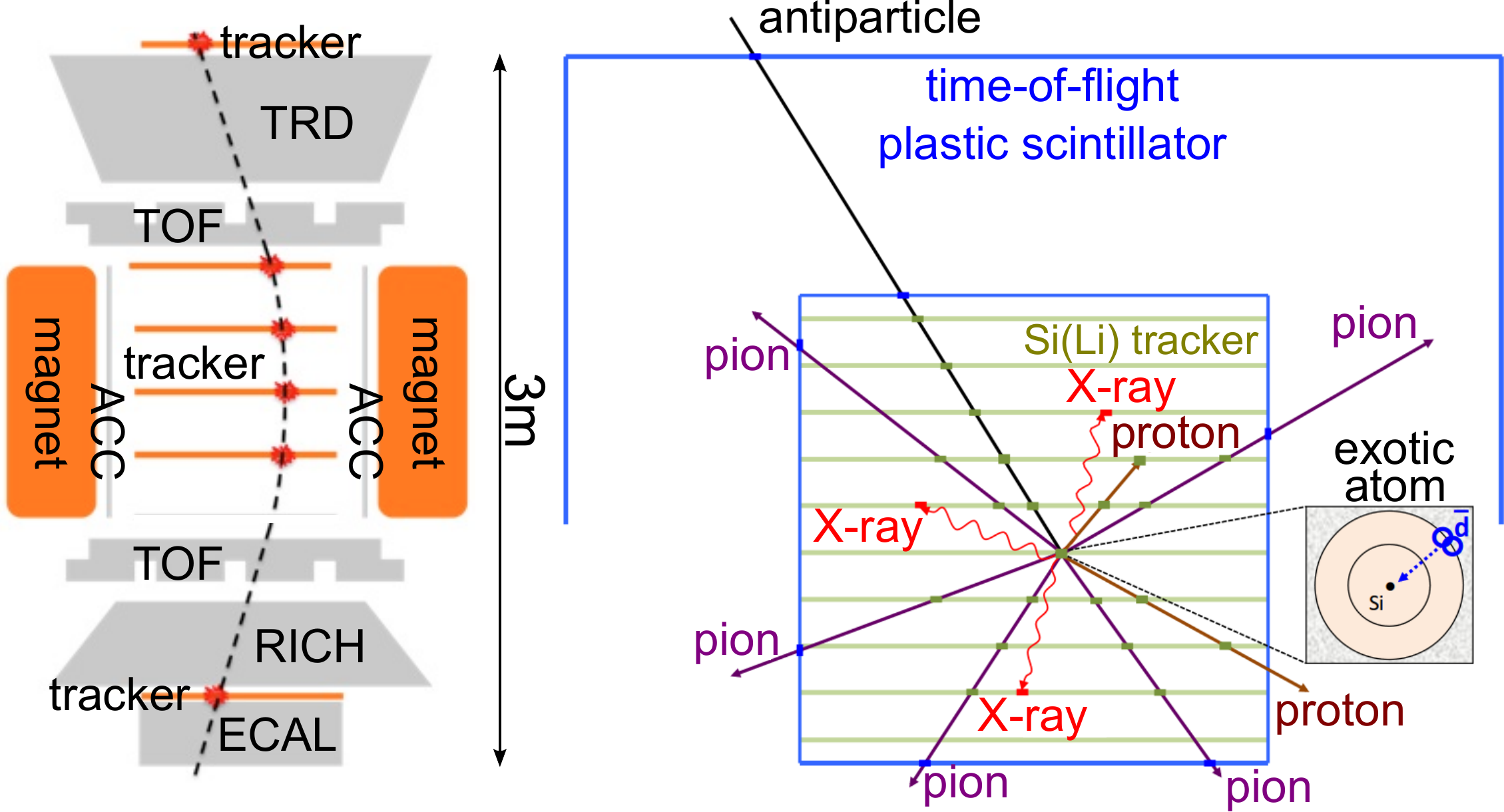}
\caption{\label{f-ams}Layout of the AMS-02 (left) and GAPS (right) detectors.}
\end{figure}

The absolute flux expected for antideuterons is very low, and thus any measurement attempt needs an exceptionally strong particle identification. The best existing antideuteron limits are given by the BESS experiment~\cite{Fuke:2005it}. In the near future this search will exclusively rely on the Alpha Magnetic Spectrometer (AMS-02)~\cite{Battiston:2008zza}, a multi-purpose cosmic-ray detector on the International Space Station (ISS), and the General AntiParticle Spectrometer (GAPS)~\cite{gaps}, which is a dedicated low-energy antideuteron detector proposed to fly several times as a long duration balloon payload from Antarctica (Fig.~\ref{f-ams}). Fig.~\ref{f-dmdbar} shows that both experiments for the first reach time the sensitivity to probe the predictions of well-motivated dark matter models. AMS-02 and GAPS have mostly complementary kinetic energy ranges, but also some overlap in the interesting low-energy region, which allows both the study of a large energy range and independent cross-checks of the results. Another very important virtue from the combination of AMS-02 and GAPS comes from the different detection techniques. AMS-02 follows the principle of typical particle physics detectors, but shrunk to the size of the Space Shuttle payload bay. Particles are identified by analyzing the event signatures of different subsequent subdetectors (transition radiation detector, time-of-flight, anticoincidence counter, silicon tracker inside of a strong magnetic field, ring imaging Cherenkov counter, electromagnetic calorimeter). The GAPS detector will consist of ten planes of lithium-drifted silicon (Si(Li)) solid state detectors and a surrounding time-of-flight system. The antideuterons will be slowed down in the Si(Li) material, replace a shell electron and form an excited exotic atom. The atom will be deexcited by characteristic X-ray transitions and will end its life by annihilation with the nucleus producing a characteristic number of protons and pions. The combination of AMS-02 and GAPS allows the study of both a large energy range and independent experimental confirmation, which is crucial for a rare event search like the hunt for cosmic-ray antideuterons. This is similar to the direct dark matter searches where more than ten running experiments are compared. 

\section{Conclusions}

This summary is based on the outcomes of the ``$\bar {\text d}$14 -- 1$^{\text{st}}$ dedicated cosmic-ray antideuteron workshop'' and reviewed the status of cosmic-ray antideuteron searches for the identification of dark matter. Recent years have seen a lot of progress and antideuterons might offer a potential breakthrough in unexplored phase space. Please refer to \cite{2015arXiv150507785A} for a more complete discussion.

\section*{Acknowledgments}

The first dedicated cosmic-ray antideuteron workshop at UCLA in June 2014 was supported, in part, by the  University of California, Los Angeles. The work of MG was supported by the Forschungs- und Wissenschaftsstiftung Hamburg through the program ``Astroparticle Physics with Multiple Messengers'' and by the Marie Curie ITN ``INVISIBLES'' under grant number PITN-GA-2011-289442. The work of AI was partially supported by the DFG cluster of excellence "Origin and Structure of the Universe. KPs work was supported in part by the National Science Foundation under Award No.~1202958. SW was supported by the Studienstiftung des Deutschen Volkes and by the TUM Graduate School.

%

\begin{thebibliography}{10}

\bibitem{2015arXiv150507785A}
T.~{Aramaki} et al., {\em submitted to Physics Reports} (2015) [\href{http://arxiv.org/abs/1505.0778}{{\tt arXiv:1505.0778}}].

\bibitem{darkmatter}
K.~Freese, {\em EAS Publications
  Series} {\bf 36} (2009) 113, [\href{http://arxiv.org/abs/0812.4005}{{\tt
  arXiv:0812.4005}}].

\bibitem{ams4}
L.~Accardo et al.,  {\em Physical Review Letters} {\bf 113} (2014) 121101.

\bibitem{2015arXiv150404276G}
G.~{Giesen} et al. (2015)
  [\href{http://arxiv.org/abs/1504.0427}{{\tt arXiv:1504.0427}}].

\bibitem{2015arXiv150501236K}
K.~{Kohri} et al. (2015)
  [\href{http://arxiv.org/abs/1505.0123}{{\tt arXiv:1505.0123}}].

\bibitem{2015arXiv150405937H}
K.~{Hamaguchi} et al. (2015)
  [\href{http://arxiv.org/abs/1504.0593}{{\tt arXiv:1504.0593}}].

\bibitem{Fornengo:2013osa}
N.~Fornengo et al., {\em J. Cosm. Astropart. Phys.} {\bf 1309} (2013) 031,
  [\href{http://arxiv.org/abs/1306.4171}{{\tt arXiv:1306.4171}}].


\bibitem{Baer:2005tw}
H.~Baer and S.~Profumo, {\em J. Cosm. Astropart. Phys.} {\bf 0512}
  (2005) 008, [\href{http://arxiv.org/abs/astro-ph/0510722}{{\tt
  astro-ph/0510722}}].

\bibitem{Donato:2008yx}
F.~Donato et al., {\em Physical Review D} {\bf 78}
  (2008) 043506, [\href{http://arxiv.org/abs/0803.2640}{{\tt
  arXiv:0803.2640}}].

\bibitem{Duperray:2005si}
R.~{Duperray} et al.,
  {\em Physical Review D} {\bf 71} (2005) 083013,
  [\href{http://arxiv.org/abs/astro-ph/0503544}{{\tt astro-ph/0503544}}].

\bibitem{Ibarra:2012cc}
A.~Ibarra and S.~Wild, {\em J. Cosm. Astropart. Phys.} {\bf 1302} (2013) 021,
  [\href{http://arxiv.org/abs/1209.5539}{{\tt arXiv:1209.5539}}].

\bibitem{Fuke:2005it}
H.~{Fuke} et al.,  {\em Physical
  Review Letters} {\bf 95} (2005) 081101,
  [\href{http://arxiv.org/abs/astro-ph/0504361}{{\tt astro-ph/0504361}}].

\bibitem{Dal:2014nda}
L.~Dal and A.~Raklev, {\em Physical Review D} {\bf 89} (2014) 103504,
  [\href{http://arxiv.org/abs/1402.6259}{{\tt arXiv:1402.6259}}].

\bibitem{Ibarra2013a}
A.~Ibarra and S.~Wild, {\em Physical Review D} {\bf 88} (2013) 023014,
  [\href{http://arxiv.org/abs/1301.3820}{{\tt arXiv:1301.3820}}].

\bibitem{ams02dbaricrc2007}
V.~{Choutko} and F.~{Giovacchini}, {\em International Cosmic Ray Conference} {\bf 4}
  (2008) 765.

\bibitem{gaps}
C.~J. Hailey, {\em New Journal of Physics} {\bf 11} (2009)
  105022.

\bibitem{Aramaki2015}
T.~Aramaki et al., {\em submitted to Astroparticle Physics} (2015),
  [\href{http://arxiv.org/abs/1506.02513}{{\tt arXiv:1506.02513}}].

\bibitem{Brauninger:2009pe}
C.~B. Br\"{a}uninger and M.~Cirelli, {\em Physics Letters B} {\bf 678} (2009) 20,
  [\href{http://arxiv.org/abs/0904.1165}{{\tt arXiv:0904.1165}}].

\bibitem{Hryczuk:2014hpa}
A.~Hryczuk et al.,  {\em J. Cosm. Astropart. Phys.} {\bf 1407} (2014) 031,
  [\href{http://arxiv.org/abs/1401.6212}{{\tt arXiv:1401.6212}}].

\bibitem{Lavalle:1900wn}
J.~Lavalle et al., {\em
  Astronomy \& Astrophysics} {\bf 479} (2008) 427,
  [\href{http://arxiv.org/abs/0709.3634}{{\tt arXiv:0709.3634}}].

\bibitem{PhysRev.129.854}
A.~Schwarzschild and {\v C}.~Zupan\v{c}i\v{c}, {\em Physical Review} {\bf 129} (1963) 854.

\bibitem{Kadastik:2009ts}
M.~Kadastik et al.,  {\em Physics Letters B} {\bf
  683} (2010) 248, [\href{http://arxiv.org/abs/0908.1578}{{\tt
  arXiv:0908.1578}}].

\bibitem{Lees:2014iub}
J.~P. Lees et al., {\em
  Physical Review D} {\bf 89} (2014) 111102.

\bibitem{Barile:2014wma}
F.~Barile et al., {\em EPJ Web Conf.} {\bf 95} (2015) 04003,
  [\href{http://arxiv.org/abs/1411.1941}{{\tt arXiv:1411.1941}}].

\bibitem{na61}
N.~{Abgrall} et al.,  {\em Journal of
  Instrumentation} {\bf 9} (2014) 6005P,
  [\href{http://arxiv.org/abs/1401.4699}{{\tt arXiv:1401.4699}}].

\bibitem{1998ApJ...509..212S}
A.~W. {Strong} and I.~V. {Moskalenko}, {\em Astrophysical Journal} {\bf 509} (1998)
  212, [\href{http://arxiv.org/abs/astro-ph/9807150}{{\tt
  astro-ph/9807150}}].

\bibitem{Maurin:2001sj}
D.~Maurin et al., {\em
  Astrophysical Journal} {\bf 555} (2001) 585,
  [\href{http://arxiv.org/abs/astro-ph/0101231}{{\tt astro-ph/0101231}}].

\bibitem{2001ApJ...563..172D}
F.~{Donato} et al., {\em Astrophysical Journal} {\bf 563} (2001)
  172, [\href{http://arxiv.org/abs/astro-ph/0103150}{{\tt
  astro-ph/0103150}}].

\bibitem{Donato:2003xg}
F.~Donato et al., {\em Physical Review D} {\bf 69} (2004)
  063501, [\href{http://arxiv.org/abs/astro-ph/0306207}{{\tt
  astro-ph/0306207}}].

\bibitem{2014PhRvD..90h1301L}
J.~Lavalle et al., {\em Physical Review D} {\bf 90} (2014) 081301.

\bibitem{gleeson-1967}
L.~J. {Gleeson} and W.~I. {Axford}, {\em Astrophysical Journal} {\bf 149} (1967) L115.

\bibitem{Maccione:2012cu}
L.~Maccione, {\em Physical Review Letters} {\bf
  110} (2013), 081101, [\href{http://arxiv.org/abs/1211.6905}{{\tt
  arXiv:1211.6905}}].

\bibitem{JGRA:JGRA16156}
N.~A. Tsyganenko, {\em Journal of
  Geophysical Research: Space Physics} {\bf 107} (2002),SMP 10--1.

\bibitem{Galoyan:2012bh}
A.~Galoyan and V.~Uzhinsky, {\em Hyperfine Interaction} {\bf 215} (2013), 69, [\href{http://arxiv.org/abs/1208.3614}{{\tt arXiv:1208.3614}}].

\bibitem{Battiston:2008zza}
R.~Battiston et al., {\em Nuclear
  Instruments and Methods in Physics Research A} {\bf 588} (2008) 227.

\end{thebibliography}

\providecommand{\href}[2]{#2}\begingroup\raggedright\endgroup

\end{document}